\documentstyle[preprint,aps]{revtex}
\begin{document}
\draft
\title{Exact Results for Hamiltonian Walks from the Solution of the
Fully Packed Loop Model on the Honeycomb Lattice}
\author{M. T. Batchelor$^a$, J. Suzuki$^b$ and C. M. Yung$^a$}
\address{$^a$Department of Mathematics, School of Mathematical Sciences,
Australian National University, Canberra ACT 0200, Australia}
\address{$^b$Institute of Physics, College of Arts and Sciences,
University of Tokyo, Komaba, Meguroku, Tokyo 153, Japan}
\date{1 June 1994}
\maketitle
\begin{abstract}
We derive the nested Bethe Ansatz solution of the fully packed
O($n$) loop model on the honeycomb lattice. From this solution
we derive the bulk free energy per site along with the central charge
and geometric scaling dimensions describing the critical behaviour. In the
$n=0$ limit we obtain the exact compact exponents $\gamma=1$ and
$\nu=1/2$ for Hamiltonian walks, along with the exact value
$\kappa^2 = 3 \sqrt 3 /4$ for the connective constant (entropy).
Although having sets of scaling dimensions in common, our results
indicate that Hamiltonian walks on the honeycomb and Manhattan lattices
lie in different universality classes.
\end{abstract}

\narrowtext

The configurational statistics of polymer chains have long been
modelled by self-avoiding walks. In the low-temperature limit
the enumeration of a single self-attracting polymer in dilute
solution reduces to that of compact self-avoiding walks.
A closely related problem isi that of
Hamiltonian walks in which the self-avoiding walk visits each site
of a given lattice and thus completely fills the available space.
Hamiltonian walks are directly related to the
Gibbs-DiMarzio theory for the glass transition of polymer melts\cite{gkm}.
More than thirty years ago now Kasteleyn obtained the exact number
of Hamiltonian walks on the Manhattan oriented square lattice\cite{k}.
More recently this work has been significantly extended to yield
exactly solved models of polymer melts\cite{dd}.
The critical behaviour of Hamiltonian walks
on the Manhattan lattice has also been obtained
from the $Q=0$ limit of the $Q$-state Potts model \cite{d}.
In particular this Hamiltonian walk problem has been shown\cite{dd,d}
to lie in the same universality class as dense self-avoiding walks,
which follow from the $n=0$ limit in the low-temperature or densely
packed phase of the honeycomb O($n$) model\cite{n,dense}.

As exact results for Hamiltonian walks are confined to the Manhattan
lattice, the behaviour of Hamiltonian walks on non-oriented lattices,
and the precise scaling of compact two-dimensional
polymers, remains unclear\cite{opb,c1,c2}.
The exact value of the (Hamiltonian) geometric exponent $\gamma^H$ was
conjectured to be $\gamma^H=\gamma^D$, where $\gamma^D=\frac{19}{16}$
was extracted via the Coulmb gas method
for dense self-avoiding walks\cite{dense,c1}.
However, recent numerical investigations of the collapsed and compact
problems are more suggestive of the value $\gamma^H=1$\cite{c2,ct,bbgop}.

More recently, Bl{\"o}te and Nienhuis\cite{bn2} have argued that a
universality class different to dense walks governs the O($n$) model
in the zero temperature limit (the fully packed loop model).
Based on numerical evidence obtained via
finite-size scaling and transfer matrix techniques, along with
a graphical mapping at $n=1$,
they argued that the model lies in a new universality class characterized by
the superposition of a low-temperature O($n$) phase and a solid-on-solid
model at a temperature independent of $n$.
This model is identical to the Hamiltonian walk problem in the limit $n=0$.
In this Letter we present exact results for Hamiltonian walks on the
honeycomb lattice from an exact solution of this fully packed loop model.
We derive the physical quantities which characterize
Hamiltonian walks on the honeycomb lattice. These include
a closed form expression for the connectivity, or entropy,
and an exact infinite set of geometric scaling dimensions which
include a conjectured value by Bl{\"o}te and Nienhuis\cite{bn2}.
Our results settle the abovementioned controversy in favour of the
universal value $\gamma^H=1$.

In general the partition function of the O($n$) loop model
can be written as
\begin{equation}
{\cal Z}_{{\rm O}(n)} = \sum t^{{\cal N}-{\cal N}_b} n^{{\cal N}_L},
\end{equation}
where the sum is over all configurations of closed and nonintersecting loops
covering ${\cal N}_b$ bonds of the honeycomb lattice and ${\cal N}$ is the
total number of lattice sites (vertices).
Here the variable $t$ plays the role of the O($n$) temperature, $n$ is the
fugacity of a closed loop and ${\cal N}_L$ is the total
number of loops in a given configuration.

For the particular choice $t = t_c$, where\cite{n}
\begin{equation}
t_c^2 = 2 \pm \sqrt{2-n},
\end{equation}
the related vertex model is exactly solvable with a Bethe Ansatz type
solution for both periodic\cite{b1,bb,s1} and open\cite{bs} boundary
conditions. This critical line is depicted as a function of $n$ in Fig. 1.
Here we extend the exact solution curve along the line
$t=0$, where the only nonzero contributions in the partition sum (1) are for
configurations in which each lattice site is visited by a loop,
i.e. with ${\cal N}={\cal N}_b$\cite{noteb}.
This is the fully packed model
recently investigated by Bl\"ote and Nienhuis\cite{bn2}.

We consider a lattice of ${\cal N} =  2 M N$ sites
as depicted in Fig. 2, i.e. with periodic boundaries
across a finite strip of width $N$.
The allowed arrow configurations and the corresponding weights of the related
vertex model are shown in Fig. 3. Here the parameter
$n = s + s^{-1} =  2 \cos \lambda$. In Fig. 2 we also show a seam to
ensure that loops which wrap around the strip pick up the correct
weight $n$ in the partition function. The corresponding weights
along the seam are also given in Fig. 3\cite{note1}. We find that
the eigenvalues of the row-to-row transfer matrix of the vertex model are
given by
\widetext
\begin{equation}
\Lambda = \prod_{\alpha=1}^{r_1} -
{\sinh (\theta_{\alpha} - {\rm i} \frac{\lambda}{2}) \over
\sinh (\theta_{\alpha} + {\rm i} \frac{\lambda}{2})}
\prod_{\mu=1}^{r_2} - {\sinh (\phi_{\mu} + {\rm i} \lambda) \over
\sinh \phi_{\mu}}
+
{\rm e}^{{\rm i} \epsilon}
\prod_{\mu=1}^{r_2} - {\sinh (\phi_{\mu} - {\rm i} \lambda) \over
\sinh \phi_{\mu}}
\end{equation}
where the roots $\theta_{\alpha}$ and $\phi_{\mu}$ follow from
\begin{equation}
{\rm e}^{{\rm i} \epsilon} \left[ -
{\sinh (\theta_{\alpha} - {\rm i} \frac{\lambda}{2}) \over
\sinh (\theta_{\alpha} + {\rm i} \frac{\lambda}{2})} \right]^N = -
\prod_{\mu=1}^{r_2} -
{\sinh (\theta_{\alpha} - \phi_{\mu}+{\rm i} \frac{\lambda}{2}) \over
\sinh (\theta_{\alpha} - \phi_{\mu} - {\rm i} \frac{\lambda}{2})}
\prod_{\beta=1}^{r_1}
{\sinh (\theta_{\alpha} - \theta_{\beta}-{\rm i} \lambda) \over
 \sinh (\theta_{\alpha} - \theta_{\beta}+{\rm i} \lambda)},
\quad \alpha=1,\ldots,r_1.
\end{equation}
\begin{equation}
{\rm e}^{{\rm i} \epsilon} \prod_{\alpha=1}^{r_1} -
{\sinh (\phi_{\mu}-\theta_{\alpha} - {\rm i} \frac{\lambda}{2}) \over
\sinh (\phi_{\mu}-\theta_{\alpha} + {\rm i} \frac{\lambda}{2})} = -
\prod_{\nu=1}^{r_2}
{\sinh (\phi_{\mu} - \phi_{\nu}-{\rm i} \lambda) \over
 \sinh (\phi_{\mu} - \phi_{\nu}+{\rm i} \lambda)},
\quad \mu=1,\ldots,r_2.
\end{equation}
\narrowtext
Here the seam parameter $\epsilon=\lambda$ for the largest sector
and $\epsilon=0$ otherwise. Apart from the seam, this exact solution
on the honeycomb lattice follows from earlier work by Baxter on
the colourings of the hexagonal lattice \cite{b2}.
Baxter derived the Bethe Ansatz solution and evaluated the bulk partition
function per site in the region $n\ge2$. The corresponding vertex
model was later considered in the region
$n<2$ with regard to the polymer melting transition
at $n=0$\cite{si}.

More generally, the fully packed loop model can be seen to follow
from the honeycomb limit of the solvable square lattice $A_2^{(1)}$
loop model \cite{wn,r}. Equivalently, the related vertex
model on the honeycomb lattice is obtained in the appropriate
limit of the $A_2^{(1)}$ vertex model on the square lattice in the
ferromagnetic regime.
This latter model is the $su(3)$ vertex model\cite{su3}.
One can verify that the above results follow from the honeycomb limit of
the Algebraic Bethe Ansatz solution of the $su(3)$ model\cite{bvv}
with appropriate seam.
It should be noted that Reshetikhin\cite{r} has performed similar
calculations to those presented here, although in the
absence of the seam, which plays a crucial role in the underlying
critical behaviour.

Defining the finite-size free energy as $f_N = N^{-1} \log \Lambda_0$, we
derive the bulk value to be
\begin{equation}
f_\infty = \int_{-\infty}^{\infty}
{\sinh^2\! \lambda x \, \sinh (\pi -\lambda)x
\over x\, \sinh \pi x \, \sinh 3 \lambda x } dx .
\label{fbulk}
\end{equation}
This result is valid in the region $0 < \lambda \le \pi/2$, where the
Bethe Ansatz roots defining the largest eigenvalue $\Lambda_0$ are all real.
We note that the most natural system size $N$ is a multiple of 3, for which
the largest eigenvalue occurs with $r_1=2N/3$ and $r_2=N/3$ roots.
In the limit $\lambda \rightarrow 0$ $f_\infty$ reduces to the known
$n=2$ value\cite{b2,b1},
\begin{equation}
f_\infty = \log \left[ \frac{3 \Gamma^2(1/3)}{4 \pi^2}\right].
\label{flim}
\end{equation}
There is however, a cusp in the free energy at $\lambda = \pi/2$.
For $\lambda > \pi/2$ the largest eigenvalue has roots
$\theta_{\alpha}$ shifted by i\,$\pi/2$. The result for $f_\infty$
is that obtained from (\ref{fbulk}) under the interchange
$\lambda \leftrightarrow \pi - \lambda$, reflecting a symmetry between the
regions $-2 \le n \le 0$ and $0 \le n \le 2$. Thus the value (\ref{flim})
holds also at $n = -2$, in agreement with the $t_c \rightarrow 0$ limiting
value\cite{b1}.

As our interest here lies primarily in the point $n = 0$ ($\lambda = \pi/2$),
we confine our attention to the region $0 \le n \le 2$. At $n = 0$,
we find that the above result for $f_\infty$ can be evaluated exactly to
give the partition sum per site, $\kappa$, as
\begin{equation}
\kappa^2 = 3 \sqrt 3/4,
\end{equation}
and thus $\kappa =  1.13975 \ldots$ follows as the exact value
for the entropy or connective constant of Hamiltonian walks on the
honeycomb lattice. This numerical value has been obtained previously
via the same route in terms of an infinite sum \cite{s2}. Our exact result
(8) is to be compared with the open self-avoiding walk, for which
$\mu^2 = 2 + \sqrt 2$\cite{n}, and so $\mu = 1.84775 \ldots$
It follows that for self-avoiding walks on the honeycomb lattice
the entropy loss per step due to compactness, relative to the freedom
of open configurations, is exactly given by
\begin{equation}
\frac{1}{2} \log \left[ \frac{3 \sqrt 3}{4 (2+\sqrt 2)} \right]
= - 0.483161 \ldots
\end{equation}

The central charge $c$ and scaling dimensions $X_i$ defining the critical
behaviour of the model follow from the dominant finite-size corrections to the
transfer matrix eigenvalues\cite{cx}. For the central charge,
\begin{equation}
f_N \simeq f_{\infty} + \frac{\pi \zeta c}{6 N^2}.
\end{equation}
The scaling dimensions are related to the inverse correlation lengths
via
\begin{equation}
\xi_i^{-1} = \log ( \Lambda_0/\Lambda_i) \simeq 2 \pi \zeta X_i/N.
\end{equation}
Here $\zeta=\sqrt 3 /2$ is a lattice-dependent scale factor.

The derivation of the dominant finite-size corrections via the
Bethe Ansatz solution of the vertex model
follows that given for the $su(3)$ model in the
antiferromagnetic regime\cite{devega} (see, also \cite{am}).
The derivation is straightforward though tedious and we omit the details.
In the absence of the seam, we find that the central charge is $c=2$ with
scaling dimensions $X = \Delta^{(+)} + \Delta^{(-)}$, where
\begin{equation}
\Delta^{(\pm)} = \frac{1}{8}\, g \,
\mbox{\boldmath $n$}^T C\, \mbox{\boldmath $n$}
+ \frac{1}{8\, g}\,
(\mbox{\boldmath $h$}^{\pm})^{T} C^{-1} \mbox{\boldmath $h$}^{\pm} -
\frac{1}{4} \mbox{\boldmath $n$}\cdot\mbox{\boldmath $h$}^{\pm},
\end{equation}
$C$ is the $su(3)$ Cartan matrix and
$\mbox{\boldmath $n$}=(n_1,n_2)$ with $n_1$ and $n_2$ related to the number
of Bethe Ansatz roots via $r_1 = 2N/3 - n_1$ and $r_2 = N/3 - n_2$\cite{note2}.
The remaining parameters $\mbox{\boldmath $h$}^{\pm} = (h_1^{\pm},h_2^{\pm})$
define the
number of holes in the root distribution in the usual way\cite{devega}.
We have further defined the variable $g = 1 - \lambda/\pi$.

With the introduction of the seam $\epsilon = \lambda$, we find that the
central charge of the fully packed O($n$) model is exactly given by
\begin{equation}
c = 2 - 6(1-g)^2/g.
\end{equation}
This is the identification made by Bl\"ote and Nienhuis\cite{bn2}.
At $n=0$ we have $g=1/2$, and thus $c = -1$.
On the other hand, both Hamiltonian walks on the Manhattan lattice\cite{dd,d}
and dense self-avoiding walks\cite{dense} lie in a different universality
class with $c=-2$. However, as we shall see below, they
do share common sets of scaling dimensions and thus critical exponents.
This sharing of exponents between the fully packed and densely packed
loop models has already been anticipated by
Bl\"ote and Nienhuis in their identification of the leading thermal
and magnetic exponents\cite{bn2}. Here we derive an exact infinite
set of scaling dimensions.

Of most interest is
the so-called watermelon correlator, which measures the geometric correlation
between $L$ nonintersecting self-avoiding walks tied together at their
extremities $\mbox{\boldmath $x$}$ and $\mbox{\boldmath $y$}$.
It has a critical algebraic decay,
\begin{equation}
\langle
\phi_L(\mbox{\boldmath $x$}) \phi_L (\mbox{\boldmath $y$}) \rangle_c \sim
|\mbox{\boldmath $x$}-\mbox{\boldmath $y$}|^{-2 X_L},
\end{equation}
where $X_L$ is the scaling dimension of the conformal source operator
$\phi_L(\mbox{\boldmath $x$})$\cite{dense}. As along the line $t=t_c$,
these scaling dimensions are associated with the largest eigenvalue
in each sector of the transfer matrix.
The pertinent scaling dimensions follow from the more general result
\begin{equation}
X = \frac{1}{2}\,g\left(n_1^2+n_2^2-n_1 \, n_2 \right) -
    \frac{(1-g)^2}{2\,g}.
\end{equation}
The sectors of the transfer matrix are labelled by the Bethe Ansatz roots
via $L=n_1+n_2$. The minimum scaling dimension
in a given sector are given by $n_1=n_2=k$ for $L=2k$ and
$n_1=k-1, n_2=k$ or $n_1=k, n_2=k-1$ for $L=2k-1$. Thus we have the
set of geometric scaling dimensions $X_L$ corresponding to the operators
$\phi_L$ for the loop model,
\begin{eqnarray}
X_{2 k-1} &=& \frac{1}{2}\,g \left(k^2-k+1\right) - \frac{(1-g)^2}{2\,g},\\
X_{2 k} &=& \frac{1}{2}\,g\, k^2 - \frac{(1-g)^2}{2\,g},
\end{eqnarray}
where $k=1,2,\ldots$
The magnetic scaling dimension is given by $X_{\sigma}=X_1$ which
agrees with the identification made in \cite{bn2}. The eigenvalue
related to $X_2$ appears in the $n_d=2$ sector of the loop model,
i.e. with two dangling bonds\cite{bn2}.
At $n=0$ this more general set of dimensions reduces to
\begin{eqnarray}
X_{2 k-1} &= \frac{1}{4} \left(k^2-k\right), \\
X_{2 k} &= \frac{1}{4} \left(k^2-1\right).
\end{eqnarray}
In comparison, the scaling dimensions for dense self-avoiding walks
are\cite{dense}
\begin{equation}
X^{\rm DSAW}_L = {\mbox{\small $\frac{1}{16}$}}\left(L^2 - 4 \right).
\end{equation}
Thus we have the relations
\begin{eqnarray}
X_{2 k-1} &=& X^{\rm DSAW}_{2 k -1} + {\mbox{\small $\frac{3}{16}$}},\\
X_{2 k} &=& X^{\rm DSAW}_{2 k}.
\end{eqnarray}
Note that $X_1=X_2=0$ and $X_L > 0$ for $L>2$. Identifying
$X_{\epsilon}=X_2$ as in \cite{dense}, then the exponents
$\gamma=1$ and $\nu = 1/2$ follow in the usual way\cite{note3}.
These are indeed the exponents to be expected for compact or space filling
two-dimensional polymers.

The corresponding scaling dimensions for Hamiltonian walks on the
Manhattan lattice are as given in (19)\cite{d}. Exact Bethe Ansatz
results on this model indicate that the scaling dimensions
$X_{\sigma}=X_{\epsilon}=0$,
from which one can also deduce that $\gamma=1$ and $\nu = 1/2$\cite{bosy}.
We also expect these results to hold for Hamiltonian walks on the
square lattice.
Extending the finite-size scaling
analysis of the correlation lengths for self-avoiding walks on the
square lattice \cite{dense} down to the
zero-temperature limit $t=0$, we see a clear convergence of the central charge
and leading scaling dimension to the values $c=-1$ and
$X_1=0$ for even system sizes, with $X_2=0$ exactly.  These results are the
analog of the present
study on the honeycomb lattice where $N = 3k$ is most natural in terms
of the Bethe Ansatz solution.

Our results indicate that fully packed self-avoiding walks
on the honeycomb lattice have the same degree of ``solvability" as
self-avoiding walks on the honeycomb lattice. The fully packed loop
model with open boundaries is also exactly solvable\cite{yb}.
The derivation of the surface critical
behaviour of Hamiltonian walks is currently in progress.

\acknowledgments

It is a pleasure to thank A. L. Owczarek, R. J. Baxter, H. W. J. Bl\"ote,
B. Nienhuis and K. A. Seaton for helpful discussions and correspondence.
This work has been supported by the Australian Research Council.

\begin{figure}
\caption{The exact solution curve $t=t_c$ (2) of the O($n$) model on the
honeycomb lattice as a function of $n$. In this Letter we extend the exact
solution curve along the line $t=0$.}
\end{figure}

\begin{figure}
\caption{The periodic honeycomb lattice of width $L$. Dashed lines
indicate the position of the seam.}
\end{figure}

\begin{figure}
\caption{(a) The allowed arrow configurations and their corresponding
vertex weights. (b) The modified vertex weights along the seam.}
\end{figure}


\begin{references}

\bibitem{gkm} See, e.g., M. Gordon, P. Kapadia and A. Malakis,
J. Phys. A {\bf 9}, 751 (1976); J. F. Nagle, P. D. Gujrati and
M. Goldstein, J. Phys. Chem. {\bf 74}, 2596 (1984);
T. G. Schmalz, G. E. Hite and D. J. Klein, J. Phys. A {\bf 17}, 445 (1984);
H. S. Chan and K. A. Dill, Macromolecules {\bf 22}, 4559 (1989)
and references therein.
\bibitem{k} P. W. Kasteleyn, Physica {\bf 29}, 1329 (1963).
\bibitem{dd} B. Duplantier and F. David, J. Stat. Phys. {\bf 51}, 327 (1988).
\bibitem{d} B. Duplantier, J. Stat. Phys. {\bf 49}, 411 (1987).
\bibitem{n} B. Nienhuis, Phys. Rev. Lett. {\bf 49}, 1062 (1982).
\bibitem{dense} B. Duplantier, J. Phys. A {\bf 19}, L1009 (1986);
H. Saleur, Phys. Rev. B {\bf 35}, 3657 (1987);
B. Duplantier and H. Saleur, Nucl. Phys. B {\bf 290}, 291 (1987).
\bibitem{opb} A. L. Owczarek, T. Prellberg and R. Brak,
Phys. Rev. Lett. {\bf 70} 951 (1993).
\bibitem{c1} B. Duplantier, Phys. Rev. Lett. {\bf 71} 4274 (1993).
\bibitem{c2} A. L. Owczarek, T. Prellberg and R. Brak,
Phys. Rev. Lett. {\bf 71} 4275 (1993).
\bibitem{ct} C. J. Camacho and D. Thirumalai, Phys. Rev. Lett.
{\bf 71} 2505 (1993).
\bibitem{bbgop} D. Bennett-Wood, R. Brak, A. J. Guttmann, A. L. Owczarek
and T. Prellberg, J. Phys. A {\bf 27}, L1 (1994).
\bibitem{bn2} H. W. J. Bl\"ote and B. Nienhuis,
Phys. Rev. Lett. {\bf 72}, 1372 (1994).
\bibitem{b1} R. J. Baxter, J. Phys. A {\bf 19}, 2821 (1986).
\bibitem{bb} M. T. Batchelor and H. W. J. Bl\"ote, Phys. Rev. Lett.
{\bf 61}, 138 (1988); Phys. Rev. B. {\bf 39}, 2391 (1989).
\bibitem{s1} J. Suzuki, J. Phys. Soc. Jpn. {\bf 57}, 2966 (1988).
\bibitem{bs} M. T. Batchelor and J. Suzuki, J. Phys. A {\bf 26}, L729 (1993).
\bibitem{noteb} Exact information can also be obtained along the lines
$n=0$ and $n=1$. We are indebted to R. J. Baxter for this remark.
\bibitem{note1} There are several ways to define the seam. This particular
choice is consistent with the vertex weight gauge factors and the
seam used in the corresponding solution of the vertex model along the
line $t=t_c$: see, Refs. \cite{b1,bb,s1}.
\bibitem{b2} R. J. Baxter, J. Math. Phys. {\bf 11}, 784 (1970).
\bibitem{si} J. Suzuki and T. Izuyama, J. Phys. Soc. Jpn. {\bf 57}, 818 (1988).
\bibitem{wn} S. O. Warnaar and B. Nienhuis,
J. Phys. A {\bf 26}, 2301 (1993).
\bibitem{r} N. Yu. Reshetikhin, J. Phys. A {\bf 24}, 2387 (1991).
\bibitem{su3} I. V. Cherednik, Theor. Math. Phys. {\bf 47}, 225 (1981);
O. Babelon, H. J. de Vega and C. M. Viallet,
Nucl. Phys. B {\bf 190} 542 (1981).
\bibitem{bvv} O. Babelon, H. J. de Vega and C. M. Viallet,
Nucl. Phys. B {\bf 200} 266 (1982).
\bibitem{s2} J. Suzuki, J. Phys. Soc. Jpn. {\bf 57}, 687 (1988).
\bibitem{cx} H. W. J. Bl\"ote, J. L. Cardy and M. P. Nightingale,
Phys. Rev. Lett. {\bf 56}, 742 (1986); I. Affleck, Phys. Rev. Lett. {\bf 56},
746 (1986); J. L. Cardy, Nucl. Phys. B {\bf 270}, 186 (1986).
\bibitem{devega} H. J. de Vega, J. Phys. A {\bf 21}, L1089 (1988).
\bibitem{am} F. C. Alcaraz and M. J. Martins, J. Phys. A {\bf 23}, L1079
(1990).
\bibitem{note2} The parameters $n_1$ and $n_2$ are related to
de Vega's parameters $S_1$ and $S_2$ via the action of the Cartan matrix:
$S_1=2n_1-n_2$ and $S_2=2n_2-n_1$.
\bibitem{note3} Using $\eta=2 X_{\sigma}$,
$1/\nu = 2 - X_{\epsilon}$ and $\gamma = (2-\eta)\nu$.
\bibitem{bosy} M. T. Batchelor, A. L. Owczarek, K. A. Seaton and C. M. Yung,
``Surface critical behaviour of an O($n$) loop model related to two
Manhattan lattice walk problems", A.N.U preprint MRR051-94.
\bibitem{yb} C. M. Yung and M. T. Batchelor, ``Integrable vertex and loop
models on the square lattice with open boundaries via reflection
matrices", A.N.U preprint MRR042-94.

\end{references}
\end{document}